\documentstyle[psfig]{basi}
\begin{document}
\title[Radio Supernovae with GMRT]{Radio Studies of Young Core 
Collapse Supernovae
}
\author[P. Chandra et al.]%
       {Poonam Chandra$^{1,2}$, Alak Ray$^2$, Sanjay Bhatnagar$^{3,4}$ \\ 
$^1$ Joint Astronomy Programme, IISc, Bangalore 560012\\
$^2$ Tata Institute of Fundamental Research, Mumbai 400005\\
$^3$ National Centre for Radio Astrophysics, TIFR, Pune 400017 \\
$^4$ National Radio Astronomy Observatory, Soccorro, NM 87801, USA}
\maketitle
\label{firstpage}
\begin{abstract}
Multi-frequency observations of two radio bright SNe (SN1993J,
SN1979C) carried out over a year with GMRT are presented. 
Their radio light curves trace the evolution of the
mass loss in the stellar wind before the pre-supernova star exploded.
Their spectra at low frequencies distinguish between the
synchrotron self-absorption and free-free absorption based models.
Two very similar type IIp SN1999gi and SN1999em  were also
observed and we comment on their progenitor mass. 
These observations of different subclasses of core-collapse supernovae
will aid in defining their progenitors. 
\end{abstract}

\begin{keywords}
Core collapse supernovae -- SN1993J, SN1979C -- progenitors 
\end{keywords}
\section{GMRT observations of supernovae}
The ejecta from a core collapse SN interacts with the circumstellar 
medium (CSM), soon after the explosion. The SN shock (with speed $\sim$
1000 times the speed of the progenitor wind) quickly probes the 
region of the wind which was lost many thousand years before. It thus probes the past history of the star.  Relativistic electrons and 
the magnetic fields inthe interaction region give rise to the 
radio-emission. 
The shock heated SN ejecta and the CSM also emit X-rays.
\subsection{SN1993J and SN1979C}
In the next page, we show the GMRT maps of SN1993J in four separate frequency 
bands. The table gives the summary 
of GMRT observations of SN1993J and SN1979C. 
We also display the low frequency spectrum of SN1993J around
day 3100.
\begin{figure}
\vspace*{-1.5cm}
\hspace*{-3.5cm}
\psfig{figure=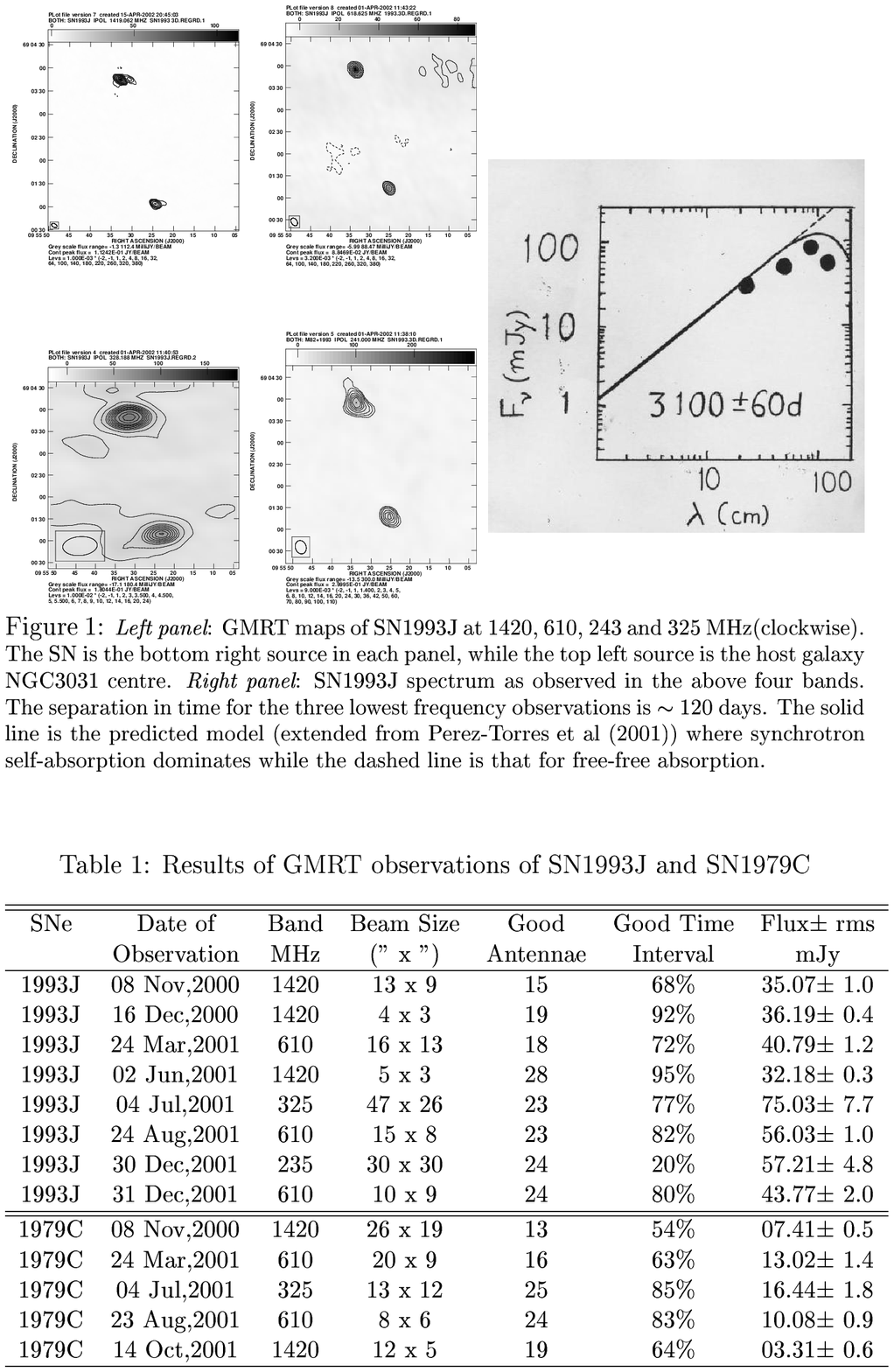,height=25.0cm,width=20.0cm,angle=0}
\end{figure}

\newpage

\subsection{Other recent supernovae}

Although radio and X-ray observations of SNe
provide complementary information, both depend upon interaction of
superova shocks with CSM.
X-ray luminosities reported by Chandra
from SN1999gi and SN1999em (both type IIP SNe) 
are relatively faint (Pooley et al 2001 and Smartt et al 2001). 
This indicates that they have low CSM density. 
Schlegel (2001) suggested that the low CSM density implied in the IIP 
SNe may support a supernova mass-loss order.
Since SN1999gi mass, as determined from its presupernova star
field's image (see Table 2 last column) is close to the 
boundary of a critical stellar mass (Chugai 1997), where it is expected
to have undergone more mass loss in the wind compared to SN1999em,
it should more radio luminous than SN1999em. This appears not to be
the case in our GMRT data (see Table 2), implying that their
CSM densities are similar. 

In the last figure we show the GMRT maps of SN2001du and SN1995N. 
SN1995N is detected by GMRT (observed on Nov 8,2000) with a flux of $\sim$ 4.5 mJy at 1420 MHz.
Finally, the table shows the supernovae that were not detected by GMRT.

\section{Conclusions}

Synchrotron self absorption of radio emission from SN1993J 
may be the dominant absorption mechanism which is determining the 
spectrum and light curve at current epochs at long wavelengths.\\ 
SN1995N is clearly detected with GMRT at 1420 MHz band. 
This Type IIn supernova is still a radio emitter and a good target for 
study of spectrum and light curves along with SN1993J and SN1979C.\\
The optically determined position of SN2001du is consistent 
(within astrometric errors and size of the synthesized beam)
with a region of enhanced radio emission at 2 mJy level at 610 MHz.

We thank the staff of the GMRT that made these observations possible. 
GMRT is run by the National Centre for Radio Astrophysics of
the Tata Institute of Fundamental Research.

\newpage
\begin{center}
\vspace*{-10cm}
Table 2. Results of GMRT observations of SN1999gi and SN1999em
\end{center}
\begin{table*}
\hspace*{1.5cm}
\begin{tabular}{ccccc}
\\
\hline
\hline
SNe & Date of     &Wave Band & Flux & Progenitor mass\\

    & Observation  &  MHz    & mJy  & (Smartt et al.)\\
    \hline
    1999gi & $\sim$750 & 610     &$\le$0.7 & 9$\pm$ 3 $M_{\odot}$ \\
    1999em & $\sim$800 & 610     &$\le$1.0 & 12$\pm$ 1 $M_{\odot}$ \\

\hline
\end{tabular}
\end{table*}

\begin{figure}
\centerline{
\psfig{figure=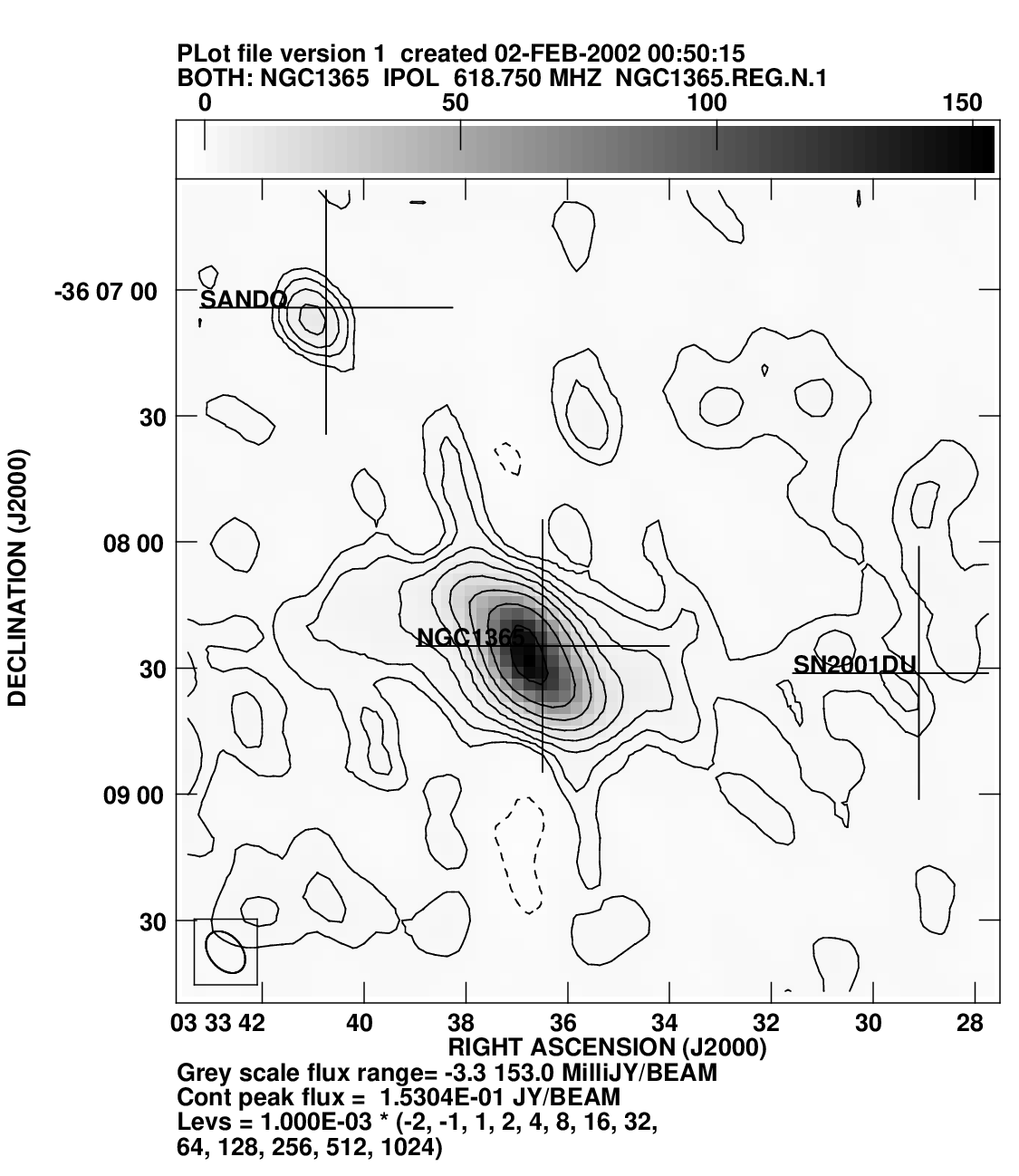,width=4.0cm}
\psfig{figure=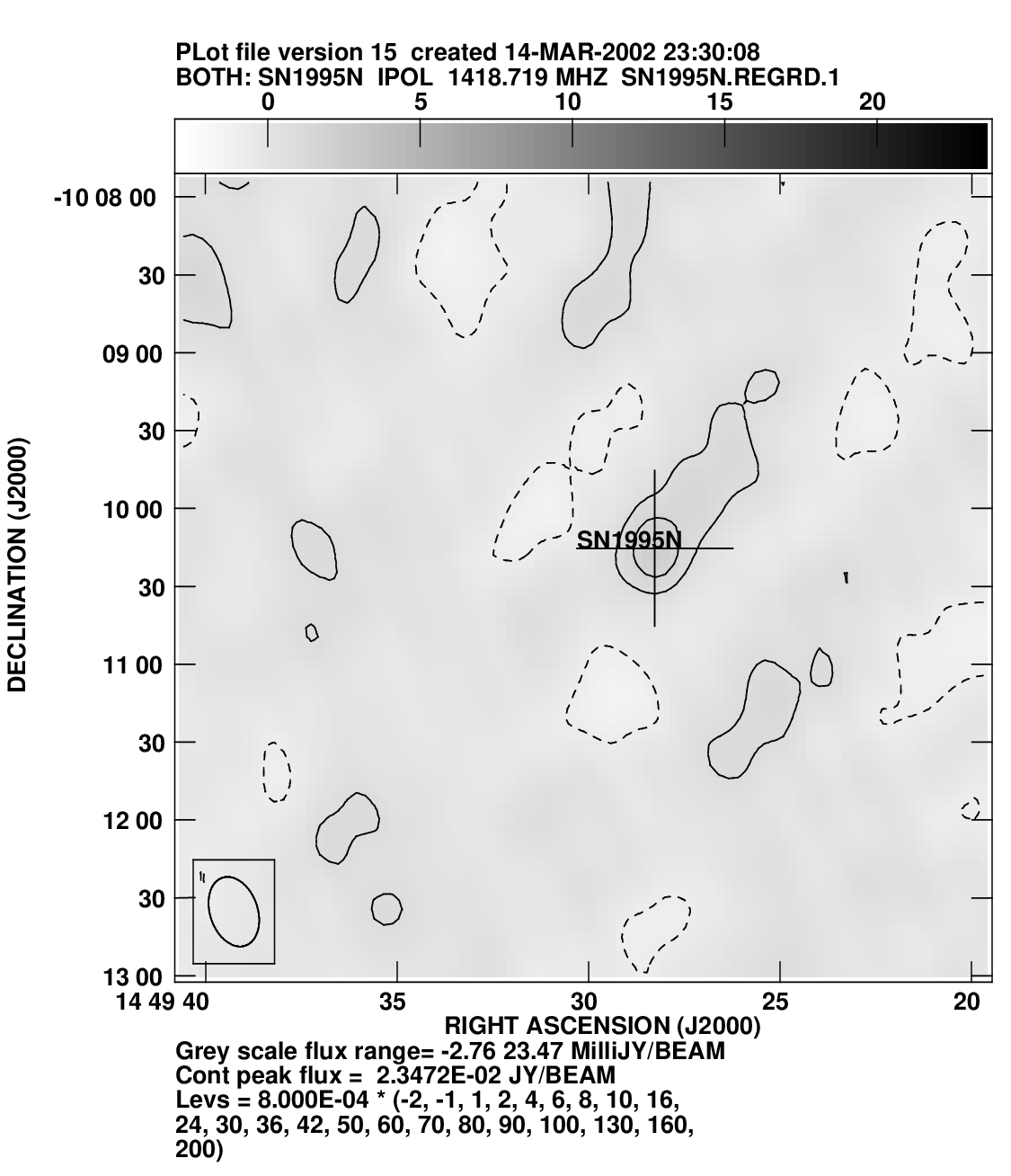,width=4.0cm}}
Figure 2. {\it Left panel}:GMRT map of
SN2001du at 610 MHz band. SN is in the right center region
near 2 mJy contour. {\it Right panel}: GMRT map of SN1995N
at 1420 MHz band. Supernova (4.5 mJy) is at the position of the cross.

\end{figure}
\begin{table*}[b]
\hspace*{0.2cm}                                             
\begin{tabular}{ccccccc}
\\
\hline
\hline
SNe & Gal.Center  &Wave Band &  Beam Size    & Good    & Good Time & Flux limit
\\

     & Offset      &  MHz     & (" x ")       &Antennae &  Interval & mJy\\
 \hline
1997ef & 10E 20S  & 1420     & 19 x 10       & 18      & 56$\%$    & 0.3\\
2001du & 90W 10S  & 1420     & 60 x 33       &  9      & 21$\%$    & 3.0\\
1983V  & 57W 30S  & 1420     & 60 x 33       &  9      & 21$\%$    & 3.0\\
1957C  & 54W 75N  & 1420     & 60 x 33       &  9      & 21$\%$    & 3.0\\
2000E  & 06W 27S  & 1420     & 13 x 7        & 20      & 63$\%$    & 0.5\\
1999el & 22E 08S  & 1420     & 13 x 7        & 20      & 63$\%$    & 0.5\\
2000W  & 02W 21S  & 1420     & 11 x 3        & 19      & 57$\%$    & 0.5\\
2000cx & 23W 109S & 1420     & 48 x 27       &  8      & 20$\%$    & 0.9\\
1999gi & 04W 61N  &  610     & 10 x 10       & 23      & 69$\%$    & 0.7\\
1999em & 15W 17S  &  610     &  8 x  6       & 24      & 73$\%$    & 1.0\\
2001ig & 139E 10N &  610     & 10 x  9       & 22      & 69$\%$    & 2.0\\
2001du & 04W 61N  &  610     & 11 x 10       & 23      & 70$\%$    & 2.0\\
\hline
\end{tabular}
\end{table*}
\begin{center}
\vspace*{9.5cm}
Table 3. Supernovae undetected by GMRT and their 2 $\sigma$ upper limits
\end{center}
\end{document}